\documentclass{desyproc}
\usepackage{graphicx}

\usepackage{bm}
\usepackage{amsmath}
\usepackage{amssymb}
\usepackage{amsfonts}
\usepackage{float}
\usepackage{dsfont}  
\usepackage{slashed}  
\usepackage{booktabs}

\newcommand{\be}{\begin{equation}}  
\newcommand{\ee}{\end{equation}}  
\newcommand{\beq}{\begin{eqnarray}}  
\newcommand{\eeq}{\end{eqnarray}}

\begin{document}
\title{Hadron structure from lattice QCD - outlook and future perspectives}

\author{{\slshape Constantia Alexandrou}\\[1ex]
Department of Physics, University of Cyprus, PO Box 20537, 1678 Nicosia, Cyprus,\\
Computation-based Science and Technology Research  
  Center, Cyprus Institute, 20 Kavafi Str., Nicosia 2121, Cyprus, \\  
  NIC, DESY, Platanenallee 6, D-15738 Zeuthen, Germany }

\contribID{307}

\confID{8648}  
\desyproc{DESY-PROC-2014-04}
\acronym{PANIC14} 

\maketitle

\begin{abstract}
We review  results on hadron structure using lattice QCD simulations with pion masses close or at the physical value. We pay particular attention to recent successes
on the computation of the mass of the low-lying baryons and on the challenges involved in evaluating energies of excited states and resonance parameters, as well as, in studies of nucleon structure.
\end{abstract}

\vspace*{-0.3cm}

\section{Introduction}
An impressive progress in algorithms and increased computational resources have allowed  lattice QCD simulations with dynamical quarks with masses fixed at  their physical values. Such simulations remove the need for a chiral
extrapolation, thereby eliminating a significant source of a systematic uncertainty that has proved difficult to quantify in the past. 
However, new challenges are presented: An increase of statistical noise  leads to large
uncertainties on most of the observables of interest. New approaches to 
deal with this problem are being developed that include better algorithms to speed up the 
computation of the quark propagators, as well as, efficient (approximate) ways to increase the statistics. 
Another challenge is related to the fact that
most of the particles become unstable if the lattice size is large enough and  methods to study decays on a finite lattice in Euclidean time need further development.

In this talk we review recent results on hadron structure  obtained using improved discretization schemes, notably  Wilson-type fermion actions and domain wall fermions. In particular, the Wilson-type twisted mass fermion (TMF) action is  particularly suitable for hadron structure studies,  mainly due to 
the automatic ${\cal O}(a)$ improvement, where $a$ is the lattice spacing. Several TMF ensembles have been produced including an ensemble simulated with two degenerate light quarks ($N_f=2$) with mass being approximately the physical value, which, for technical reasons, also includes a clover term in the action  but avoids  smearing of the gauge links~\cite{Abdel-Rehim:2013yaa}. 
 We will refer to this ensemble as the 'physical point ensemble' and  present a number of new results. The other TMF ensembles are simulated with light quarks having masses larger than physical but  where  simulations are performed for three values of $a$ allowing to study the dependence on the lattice spacing and to take the continuum limit. These ensembles include
simulations with strange and charm quarks in the sea ($N_f=2+1+1$) besides $N_f=2$ TMF ensembles. In particular, we will
use an $N_f=2+1+1$ ensemble having a pion mass $m_\pi=373$~MeV  to study lattice systematics by performing a high statistics analysis including all disconnected contributions to key nucleon observables. 

\vspace*{-0.2cm}

\section{Lattice formalism}\vspace*{-0.1cm}
An {\it ab Initio} non-perturbative solution of Quantum Chromodynamics (QCD) 
is based on defining the theory on a four-dimensional Euclidean lattice that ensures  gauge invariance. While this approach allows a direct simulation of the original theory, it introduces systematic uncertainties. These so called lattice artifacts need to be carefully investigated before lattice QCD results can be compared  to observables. In summary, in order to obtain final results in lattice QCD we need to take into account the following:

\vspace*{-0.2cm}

\begin{itemize}
\item   Due to the finite lattice spacing, simulations for at least three values of $a$  are needed in order to take the continuum limit $a\rightarrow 0 $. \vspace*{-0.2cm}
 \item Due to the finite lattice volume $L^3\times T$, simulations at different volumes are needed in order to take the infinite volume limit $L\rightarrow \infty$. For zero-temperature calculations, as the ones reported here, the temporal extent $T$ is typically twice the spatial extent $L$.\vspace*{-0.2cm}
\item  Due to the tower of QCD eigenstates entering a typical correlation function one needs a careful identification of the hadron state of interest.  How severe this so called contamination due to the  excited states  is differs depending on the observable  e.g. for the nucleon axial charge $g_A$ is found to be minimal, while for the $\sigma$-terms is large.\vspace*{-0.2cm}
\item In most hadron structure calculations contributions arising from the coupling of e.g. the electromagnetic current to sea quarks are neglected. These so called disconnected contributions are technically difficult to evaluate and have large gauge noise. They thus require new techniques and much larger  statistics as compared to the connected contributions. Taking advantage of new approaches that are particularly suited for new computer architectures such as graphic cards (GPUs) the evaluation of these diagrams to sufficient accuracy has become feasible. This has been demonstrated  for pion masses of about 300~MeV to 400~MeV~\cite{Alexandrou:2013wca,Abdel-Rehim:2013wlz,Alexandrou:2014disc,Meinel:2014}. Their applicability  for the physical point is being tested.\vspace*{-0.2cm}
\item Up to very recently, lattice QCD simulations were performed at larger than physical values of the light quark masses and thus the results
required chiral extrapolation. Simulations with light quark masses fixed to their physical values are now feasible, which eliminates a systematic error  inherent in all lattice QCD computations in the past. However, most lattice QCD results at the physical point are still preliminary since lattice artifacts have not been studied to the required accuracy. This issue is currently being addressed. 
\end{itemize}

\vspace*{-0.2cm}

In order to evaluate hadron masses one needs the computation of two-point functions. For a hadron  $h$  we construct the two-point function  of momentum {\bf p} by acting on the
vacuum with a creation operator $J^\dagger_h$ that has the quantum numbers of $h$
\beq
       \langle J_h(t_s)J_h^\dagger(0) \rangle  &=& \sum_{n,\bf {x}}e^{i{\bf p}.{\bf x}}\langle 0|J_h \>e^{-H_{QCD}t_s}|n><n|J_h^\dagger|0\rangle \\
&=&\sum_{n}|\langle 0|J_h|n \rangle|^2 e^{-E_n t_s}
\stackrel{t_s\rightarrow \infty; {\bf p}={\bf 0}} {\longrightarrow} |\langle 0|J_h|h \rangle|^2\>e^{-m_h t_s},
   \eeq   
which yields its mass for ${\bf p}={\bf 0}$ in the large Euclidean time limit.
Note that the  noise to signal increases with $ t_s$ e.g. like $e^{(m_h-\frac{3}{2}m_\pi)t_s}$ for a baryon  and thus in any lattice QCD computation there is a delicate balance between taking the large Euclidean time limit and controlling the gauge noise.
So called smearing techniques are developed that allow the construction of interpolating fields that have larger overlap with the ground state and equivalently smaller overlap with excited states so that the latter are damped out faster.

\vspace*{-0.2cm}

\section{Recent achievements}
A number of collaborations are currently producing simulations with physical values of the quark mass with each collaboration  typically using a different  ${\cal O}(a)$-improved discretization scheme. Notably, the  MILC~\cite{Bazavov:2014wgs},  BMW (Budapest-Marseille-Wuppertal)~\cite{Durr:2013goa} and ETM (European Twisted Mass)~\cite{Abdel-Rehim:2013yaa} collaborations have already generated simulations  with light quark masses fixed to their physical value using staggered, clover and twisted mass fermions, respectively. Clover gauge  configurations have also been  produced by the QCDSF~\cite{Bali:2012qs}   and PACS-CS~\cite{Aoki:2009ix} collaborations at near physical pion mass value. Recently the RBC/UKQCD collaboration reported results using domain wall fermions (DWF)  simulated with physical values of the light quark masses~\cite{Syritsyn}.
These recent developments are paving the way for lattice QCD to provide results,
which can be directly compared to experimental measurements.

\begin{figure}[h!]
\begin{minipage}{0.49\linewidth}
\includegraphics[width=\linewidth]{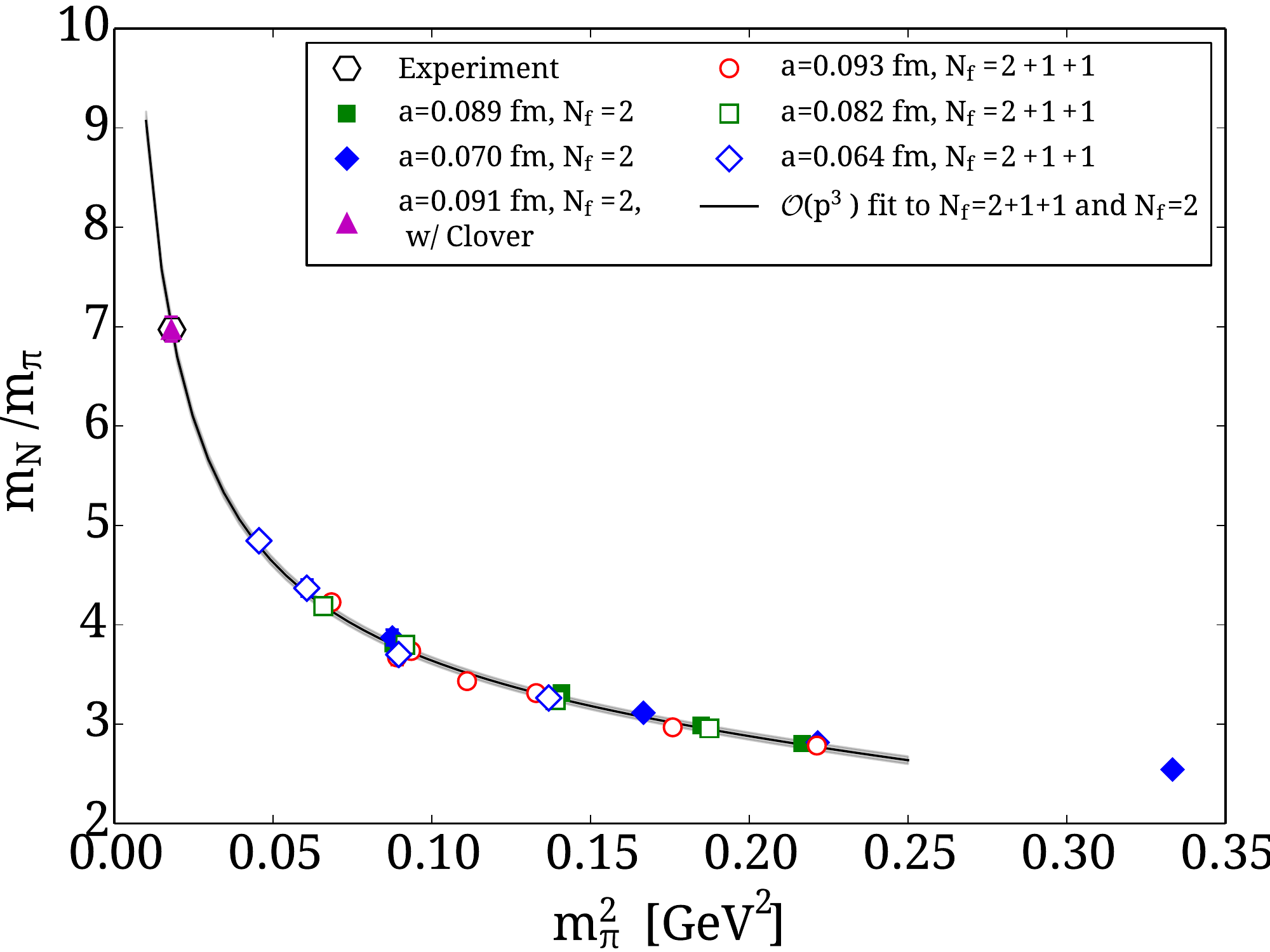}
\end{minipage}
\begin{minipage}{0.49\linewidth}
\includegraphics[width=1.1\linewidth]{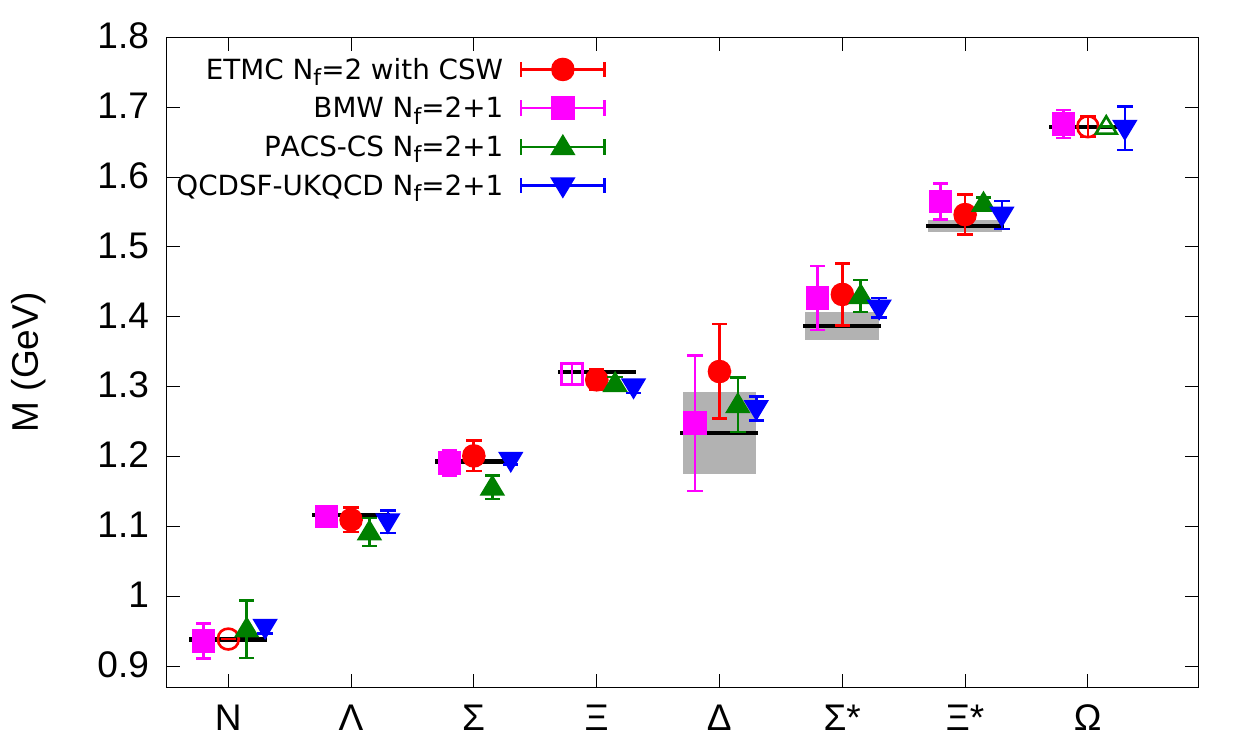}
\end{minipage}
\label{fig:mass}
\caption{Left: Results on the ratio of the proton to the pion mass versus $m_\pi^2$ using TMF. The solid line is a leading oder chiral fit using $m_\pi<300$~MeV excluding the value at the physical point. Right: Lattice QCD results on the octet and decuplet baryon masses compared to the experimental values shown by the horizontal bands.  Results by the ETM collaboration are shown in red circles
for the physical point ensemble~\cite{Alexandrou:2014}. Also shown are results using clover fermions from BMW~\cite{Durr:2008zz}  (magenta squares), from PACS-CS~\cite{Aoki:2008sm} (green triangles), and from QCDSF-UKQCD~\cite{Bietenholz:2011qq} (blue inverted triangles). Open symbols show  the baryon mass used as input to the calculations.}
\vspace*{-0.2cm}
\end{figure}

\begin{figure}[h!]
\begin{minipage}{0.49\linewidth}
\hspace*{-0.3cm}\includegraphics[width=1.1\linewidth]{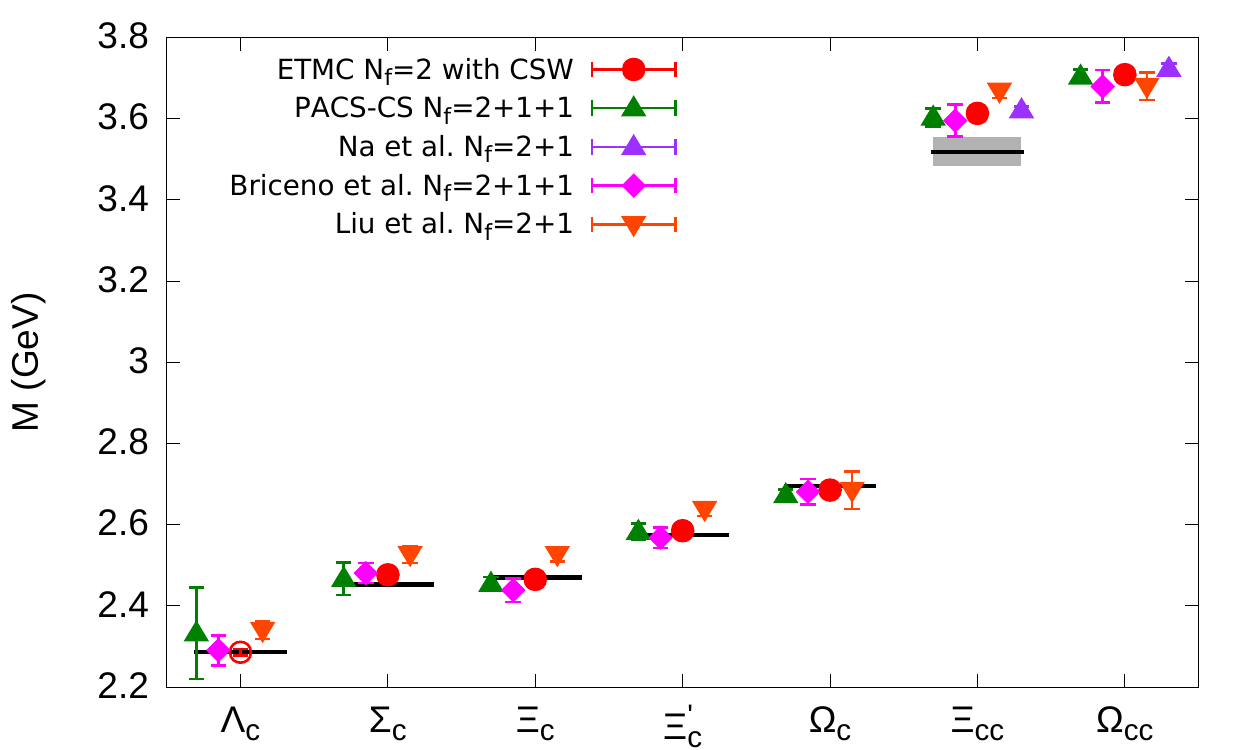}
\end{minipage}\hfill
\begin{minipage}{0.49\linewidth}
\includegraphics[width=1.1\linewidth]{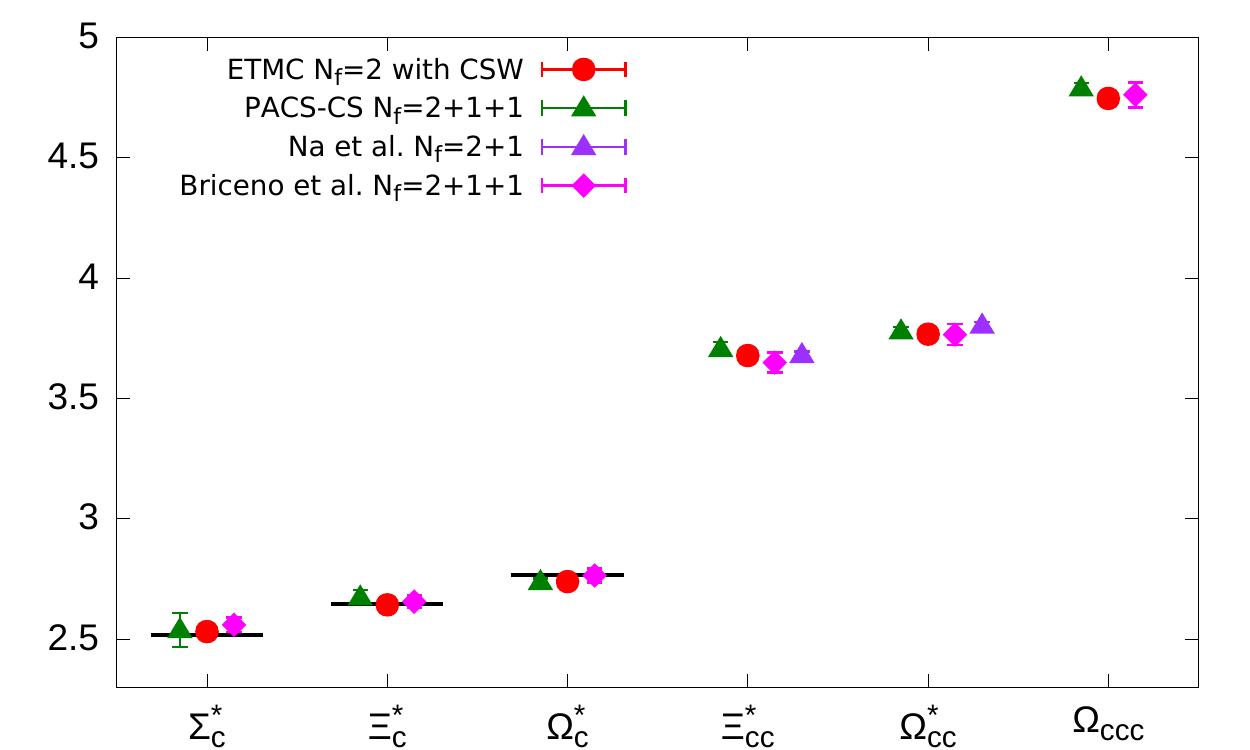}
\end{minipage}
\label{fig:mass charm}
\caption{Results by ETMC are shown in red circles for the mass of the spin-$1/2$ (left) and spin-$3/2$ (right) charmed baryons for the physical point ensemble. Included are results from various hybrid actions with staggered sea quarks from Refs.~\cite{Na:2007pv} (purple triangles), \cite{Briceno:2012wt} (magenta diamonds) and \cite{Liu:2009jc} (orange inverted triangles). Results from PACS-CS~\cite{Namekawa:2013vu} are shown in green triangles.}
\vspace*{-0.3cm}
\end{figure}

\begin{figure}[h!]
\begin{minipage}{0.49\linewidth}
\hspace*{-1.cm}\includegraphics[width=0.8\linewidth,height=1.25\linewidth,angle=-90]{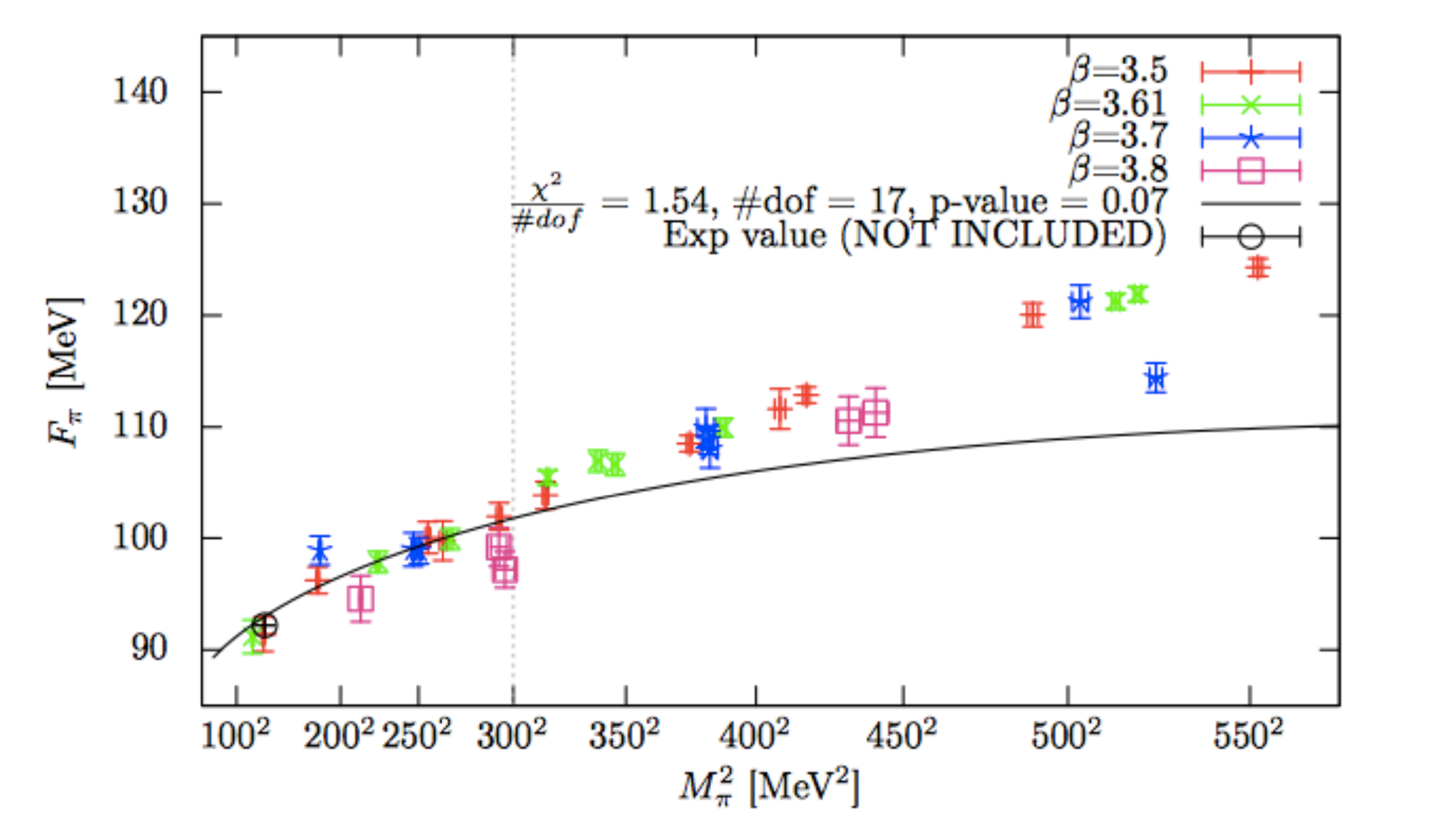}
\end{minipage}\hfill
\hspace*{0.2cm}\begin{minipage}{0.49\linewidth}\vspace*{-0.3cm}
{\includegraphics[width=1.15\linewidth]{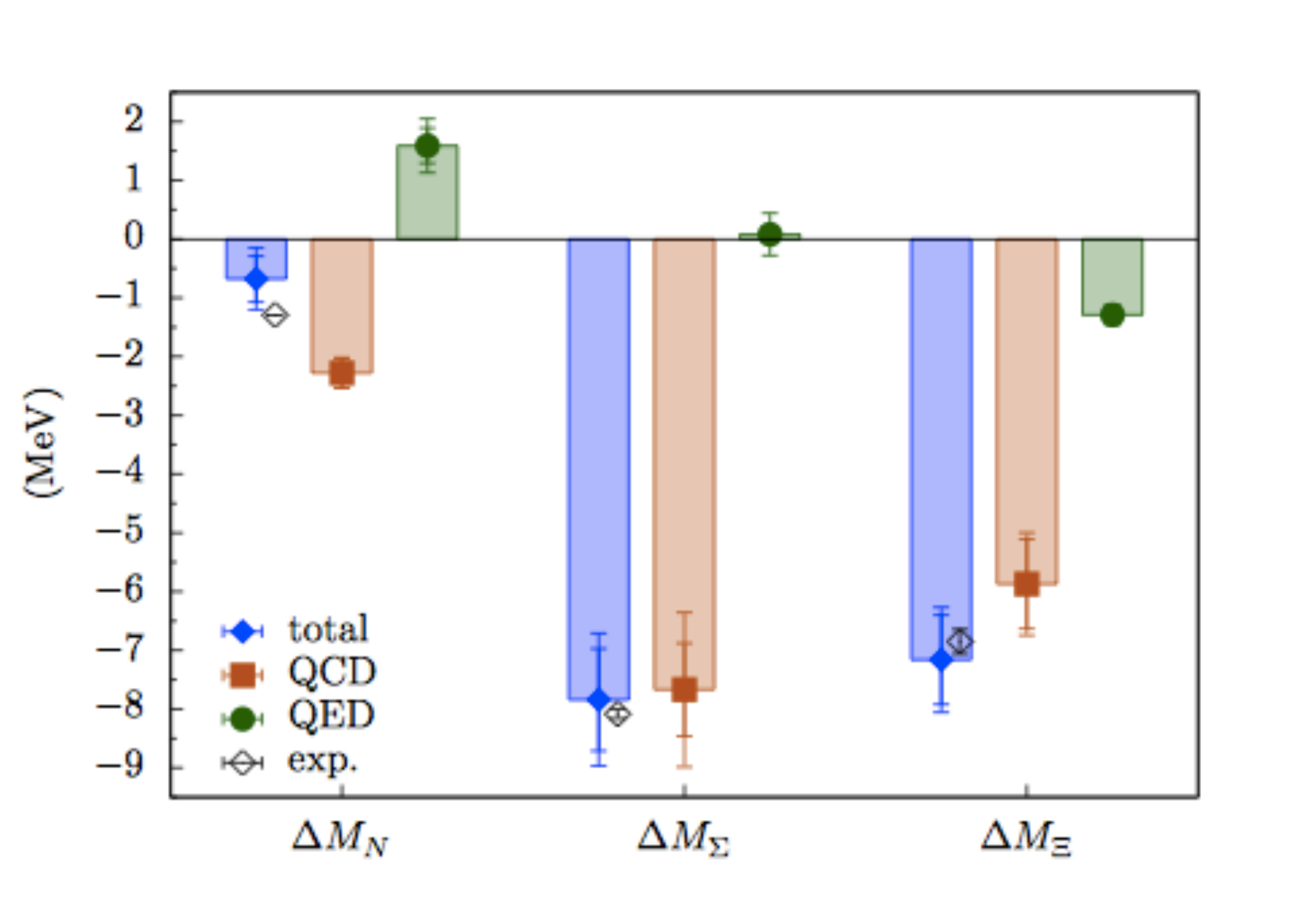}}
\end{minipage}
\caption{Left: Results on the pion decay constant $f_\pi$ by BMW. The solid line is the result of fitting NLO SU(2) chiral perturbation theory for $m_\pi<300$~MeV, taken from Ref.~\cite{Durr:2013goa}. Right: 
Baryon spectrum with mass splitting by BMW taken from Ref.~\cite{Borsanyi}.}
\label{fig:BMW}
\end{figure}

 In  Fig.~1 
 we show the ratio of the nucleon mass  to the pion mass $m_N/m_\pi$ versus $m_\pi^2$ for a number of $N_f=2$ and $N_F=2+1+1$ TMF ensembles including the
one with the physical point ensemble (with pion mass $m_\pi=130$~MeV, $a=0.094$~fm and $L=4.5$~fm) for which the dimensionless ratio $m_N/m_\pi$ agrees with its experimental value.
In the same figure we also show 
results on the low-lying baryon spectrum from the ETM, BMW, PACS-CS and QCDSF-UKQCD collaborations. 
 The  set of TMF results shown in Fig.~1 
 is obtained using the physical point  ensemble, thus requiring no chiral extrapolation, reducing drastically the systematic error that was found to be dominated by the chiral extrapolation in an earlier study using TMF~\cite{Alexandrou:2014sha}. These results are, however, obtained  at one lattice spacing and volume. The analysis of Ref.~\cite{Alexandrou:2014sha}
 has shown that lattice artifacts both due to the finite volume and lattice spacing $a$ are small and thus the values obtained for the physical point ensemble are 
expected to have small lattice artifacts.  This is indeed corroborated 
by the fact that the 'raw' lattice data  agree with the experimental values~\cite{Alexandrou:2014}.
 In Fig.~2 
 we show the corresponding results for the mass of
the charmed baryons using  the physical point ensemble in the case of TMF. As can be seen, the known values of the masses of the charmed baryons are reproduced and thus our computation
provides a prediction for the yet unmeasured masses. Our preliminary values for the
  $\Xi_{cc}^*$ is 3.678(8)~GeV, for the  $\Omega^+_{cc}$ is 3.708(10)~GeV, for $\Omega^{*+}_{cc}$  3.767(11)~GeV and for  $\Omega^{++}_{ccc}$ 4.746(3)~GeV.

The BMW collaboration has produced a number of ensembles using $N_f=2+1$ clover improved Wilson fermions with HEX smearing. They
represent the most comprehensive set of ensembles for 
light pion masses close to and at the physical point.
Their results on the pion decay constant $f_\pi$ are shown in Fig.~\ref{fig:BMW}. Fitting to NLO SU(2) chiral perturbation theory using pion masses up to 300~MeV they
reproduce the physical value of $f_\pi$. 

The BMW and QCDSF-UKQCD~\cite{Horsley:2013qka} collaborations investigated the mass splitting due to isospin breaking  and electromagnetic effects.  In Fig.~\ref{fig:BMW} we show the results on the nucleon, $\Sigma$ and $\Xi$ baryons by the BMW collaboration~\cite{Borsanyi} where isospin and electromagnetic effects were treated to lowest order. The agreement with the experimental values is a spectacular success of lattice QCD.

\section{Challenges and future perspectives}

The results shown in the previous section highlight the success of lattice QCD and the promise it holds to provide insight on many other observables.
We will briefly discuss some of the challenges that need to be addressed in order for this
to happen.
\subsection{Excited states and resonances}

In order to go beyond the low-lying spectrum one needs a formulation to extract 
excited states. The standard approach is to use a variational basis
 of interpolating fields to construct a correlation matrix of two-point functions:

\vspace*{-0.2cm}

\be
G_{jk}( t_s)= \langle{J_j(t_s)J_k^\dagger(0)} \rangle\>, j,k=1,\ldots N 
\ee
and then solve the generalized eigenvalue problem (GEVP) defined by
\vspace*{-0.2cm}
\be
 G(t)v_k(t;t_0)=\lambda_k(t;t_0) G(t_0)v_k(t;t_0) \rightarrow \lambda_k(t;t_0)=e^{-E_k(t-t_0)}\quad ,
\ee
which yields the N lowest eigenstates.

\begin{figure}[h!]\vspace*{-0.5cm}
\begin{minipage}{0.45\linewidth}
\includegraphics[width=\linewidth]{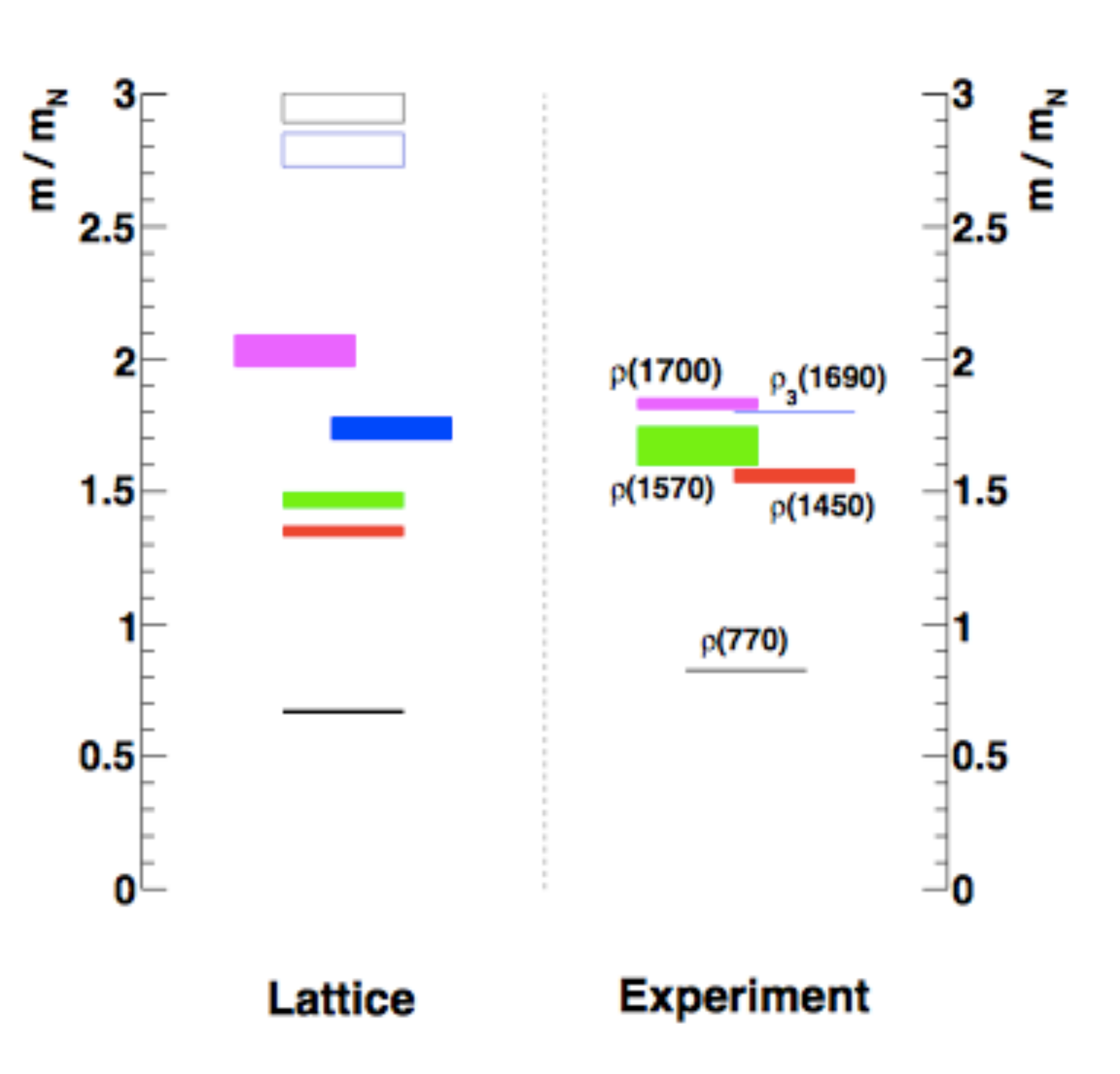}
\end{minipage}\hfill
\begin{minipage}{0.45\linewidth}
\hspace*{-1.5cm}  \includegraphics[width=0.9\linewidth,angle=-90]{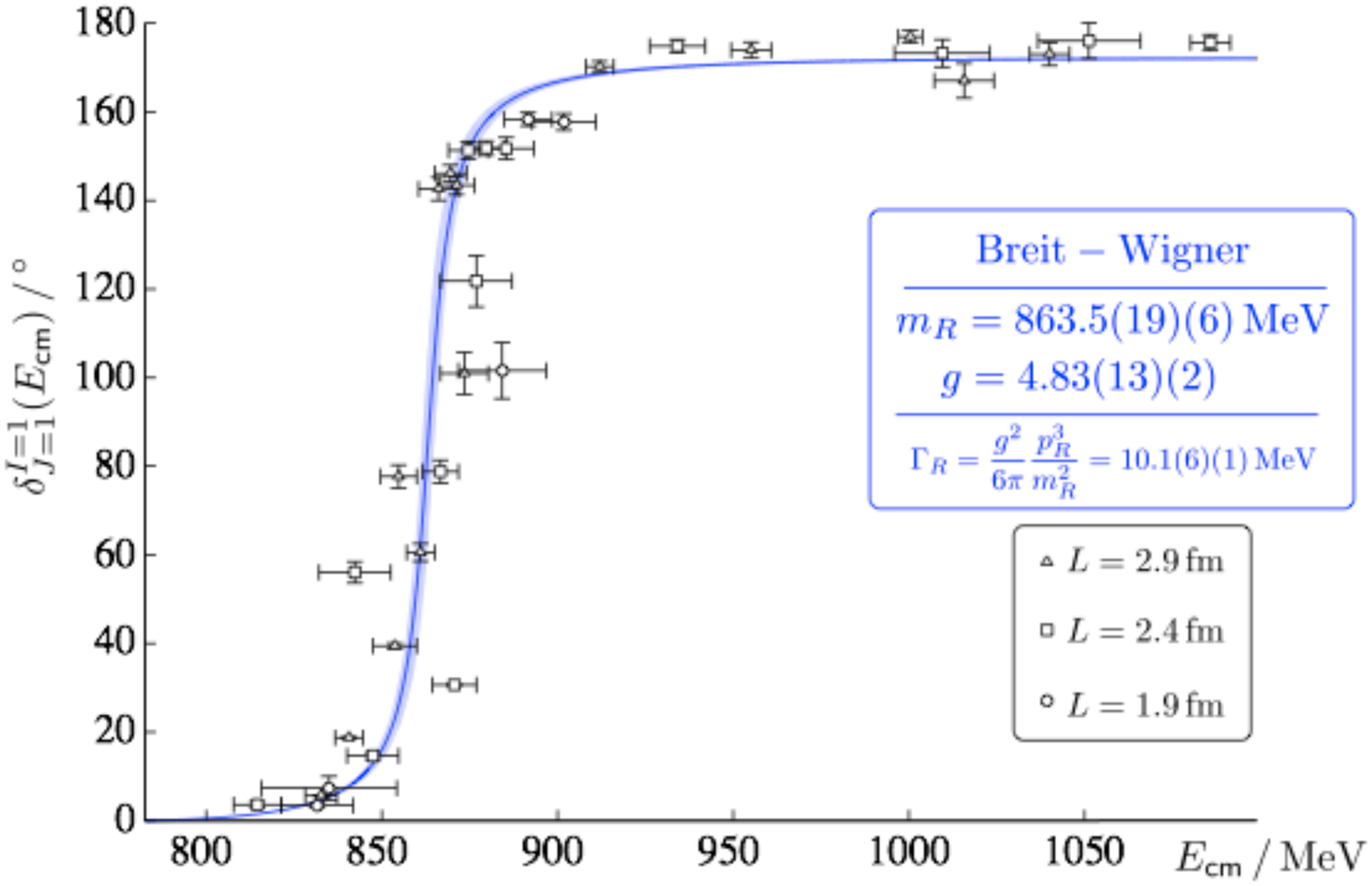} \end{minipage}
\caption{Left:
Excited states of the $\rho$-meson at $m_\pi\sim 400$~MeV at one lattice spacing and one lattice volume, compared to experiment~\cite{Bulava}. Right: The $\rho$-meson width, taken from Ref.~\cite{Dudek}.}
\label{fig:rho}
\end{figure}

A lot of effort has been devoted to construct  appropriate bases using lattice symmetries by e.g. the Hadron Spectrum Collaboration. In order to determine the
energy of an excited state one:
i) must extract all states lying below the state of interest,  ii) include disconnected diagrams,
iii) treat appropriately  resonances and unstable particles that require including  multi-hadron states.
Given the increased complexity of the problem it comes with no surprise that
the calculations performed so far have not reached the maturity of ground state 
mass computations. In Fig.~\ref{fig:rho} we show results obtained
on the $\rho$-meson  excited spectrum~\cite{Bulava} at $m_\pi=400$~MeV, as well as, on the width of the $\rho$-meson using  using $N_f=2+1$ clover fermions and 3 asymmetric lattices~\cite{Dudek}. These results, although still at larger than physical pion masses, provide a promising framework for the study of unstable particles.

\subsection{Nucleon Structure}
In order to evaluate hadron matrix elements one needs the appropriate three-point functions. There are two contributions we typically need to evaluate: the so called connected and disconnected parts, the former having the current coupled to a valence quark, while the later to a sea quark. Methods to evaluate the connected contribution are well developed see e.g.~\cite{Alexandrou:2011iu}. The disconnected contributions are much more demanding for technical reasons but also because they are prone to large gauge noise. Thus in most hadron structure computations they were neglected.

\subsubsection{Axial charges}

 Some important nucleon observables only need the connected part. These are isovector quantities for which the disconnected contributions vanish in the isospin limit. The nucleon axial charge $g_A$ is extracted from the nucleon matrix element of the isovector axial-vector current and thus it is protected from disconnected contributions. It is also well-determined experimentally from $\beta$-decays and can be extracted directly  at zero momentum transfer squared $q^2$ from $\langle N({\bf p}^\prime)|j_A| N({\bf p})\rangle|_{q^2=0}$. It thus comprises an ideal benchmark quantity for lattice QCD. 

\begin{figure}[h!]
\begin{minipage}{0.49\linewidth}
\includegraphics[width=\linewidth]{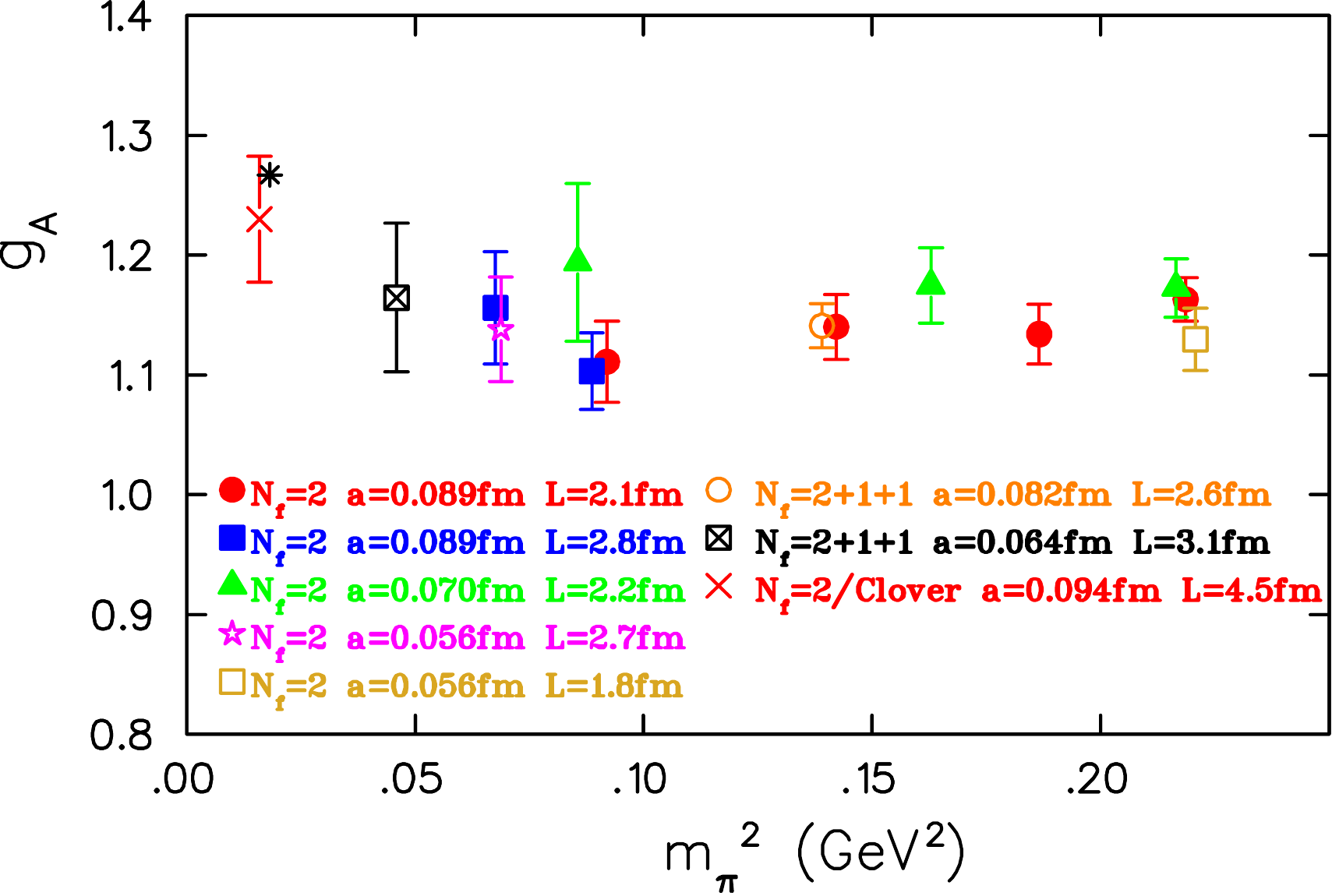}
\end{minipage}\hfill
\begin{minipage}{0.49\linewidth}
\includegraphics[width=\linewidth]{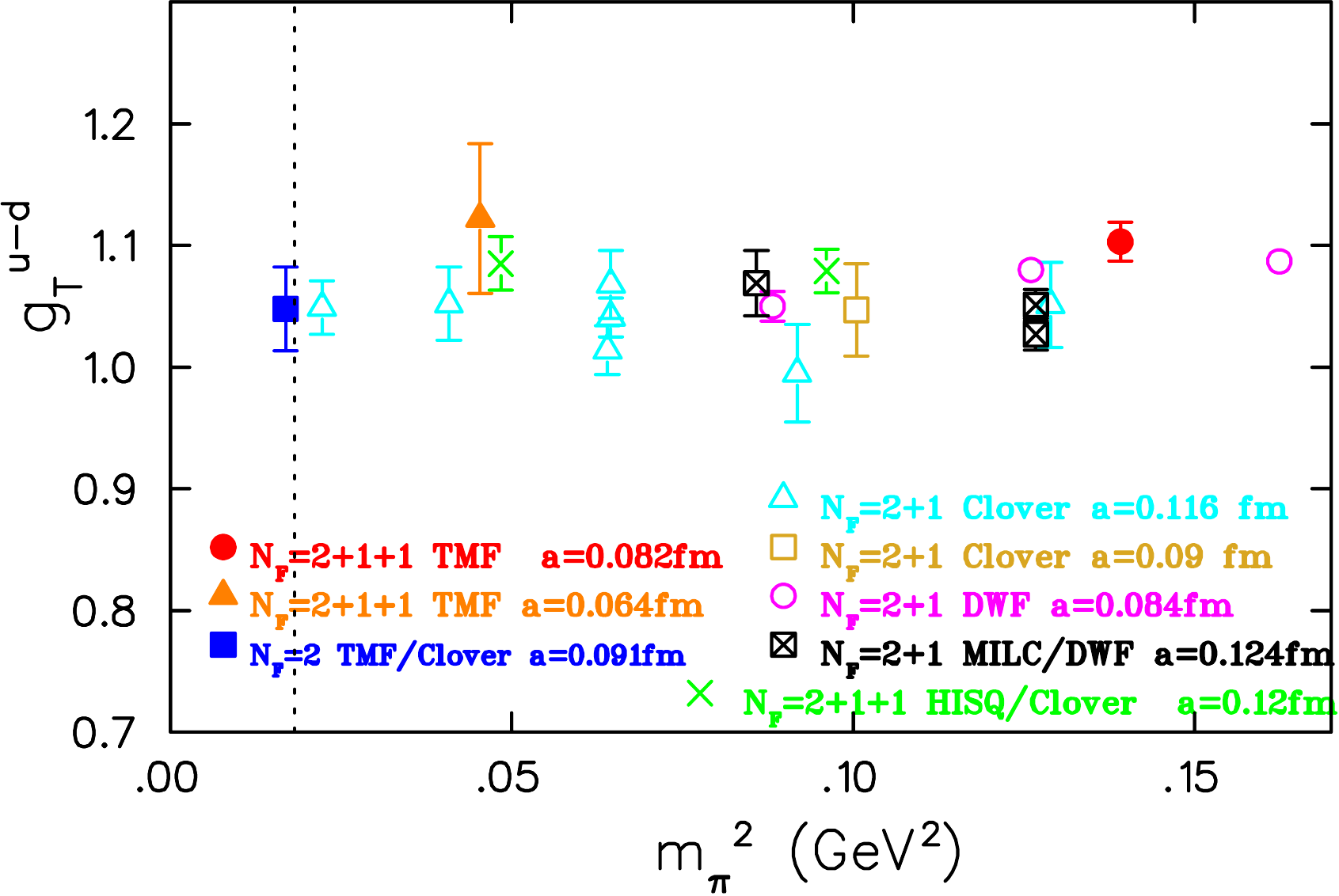}
\end{minipage}
\caption{Left: Nucleon axial charge using TMF fermions. Right: Nucleon isovector tensor charge using TMF (ETMC)~\cite{Alexandrou:2013wka}, DWF (RBC)~\cite{Aoki:2010xg}, $N_f=2$ (QCDSF-UKQCD)~\cite{Pleiter:2011gw} and $N_f=2+1$~(LHPC)~\cite{Green:2012ej} clover fermions, and  clover on $N_f=2+1+1$ staggered (PNDME)~\cite{Bhattacharya:2013ehc}. }
\label{fig:charges}
\vspace*{-0.3cm}
\end{figure}

We show in Fig.~\ref{fig:charges} results on the nucleon axial charge using the TMF ensembles. They are the 'raw' lattice QCD data in the sense that they have not been volume corrected nor extrapolated to the continuum
 limit, but  have been non-perturbatively
renormalized. They are obtained by fitting to the plateau of an appropriately defined ratio of the three- to two-functions using a sink-source separation of about 1~fm. Within the current errors no dependence on 
the lattice spacing and volume is observed. While results at higher pion mass
underestimate $g_A$, a fact observed by all lattice QCD collaborations, at the
physical point we find a value that is in agreement with experiment. Despite the fact that the statistical error is still large, this is a very welcome result that would resolve a puzzle that persisted for some time showing the importance of
computing observables at the physical point.

 Having computed the axial charge it is straight forward to calculate the isovector scalar and tensor charges. The value of the latter is particularly relevant for searching for new type of interactions beyond the Standard Model. There is a planned SIDIS on $^3$He/Proton experiment to take place at JLab after the upgrade at 11 GeV. In lattice QCD it is computed by replacing the axial-vector current by the tensor current
$j_{T}^3=\bar{\psi}(x)\sigma^{\mu\nu}\frac{\tau^3}{2}\psi(x)$. Studies have shown that $g^{u-d}_T$ has a similar behavior to $g_A$ as far as the contribution from excited states is concerned.  We show results in Fig.~\ref{fig:charges} obtained using TMF, clover, DWF and in a mixed action set-up of staggered sea and clover  valence quarks. As can seen, all lattice QCD results are in agreement and a preliminary value of $g_T^{u-d}= 1.048(34)$ in $\overline{\rm MS}$ at 2~GeV is obtained from the TMF ensemble directly at the physical point.

\subsubsection{Disconnected quark loop contributions}

\begin{figure}[h!]
\begin{minipage}{0.49\linewidth}
{\includegraphics[width=\linewidth]{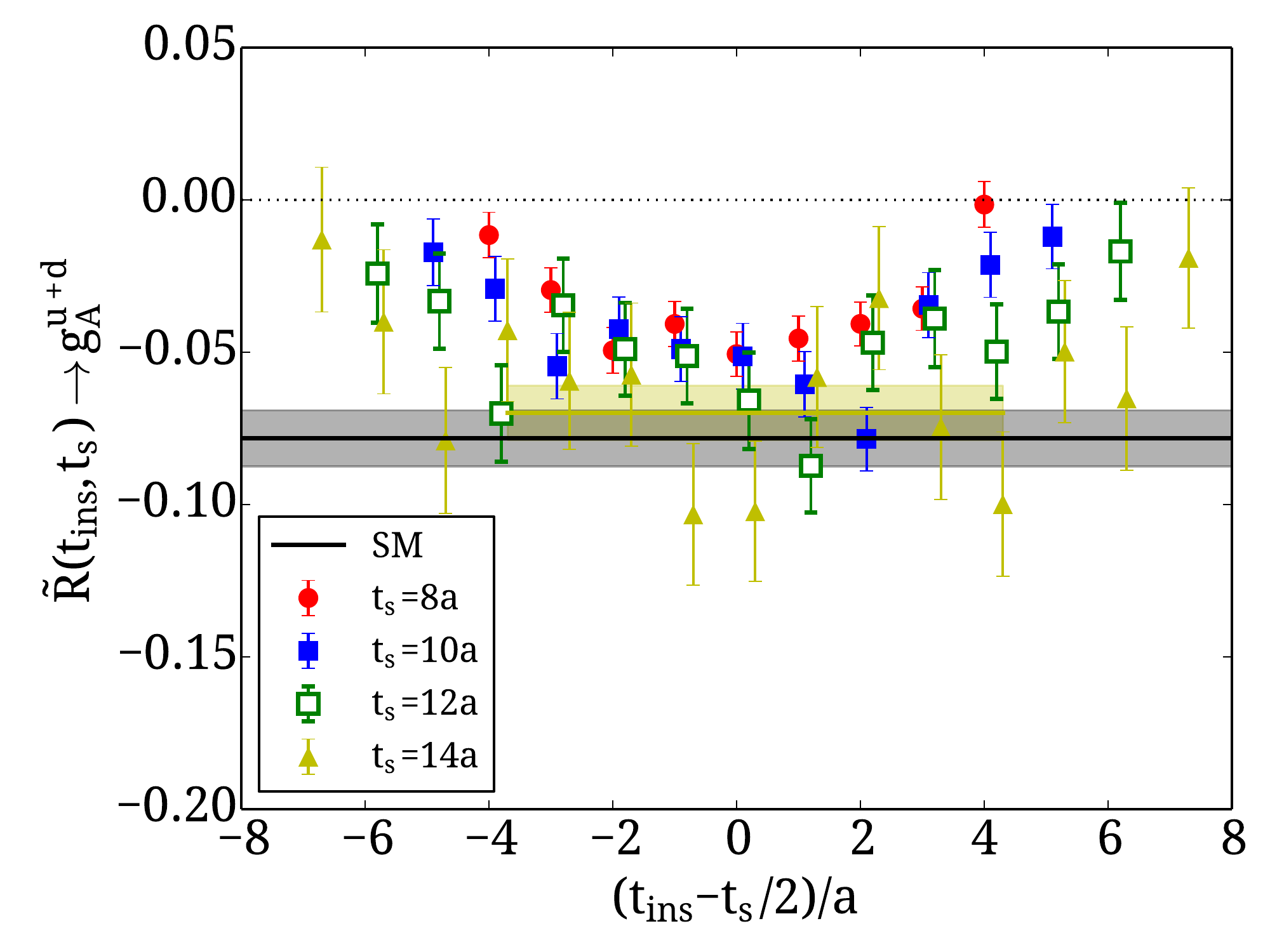}}
\end{minipage}\hfill
\begin{minipage}{0.49\linewidth} \vspace*{0cm}
{\includegraphics[width=\linewidth]{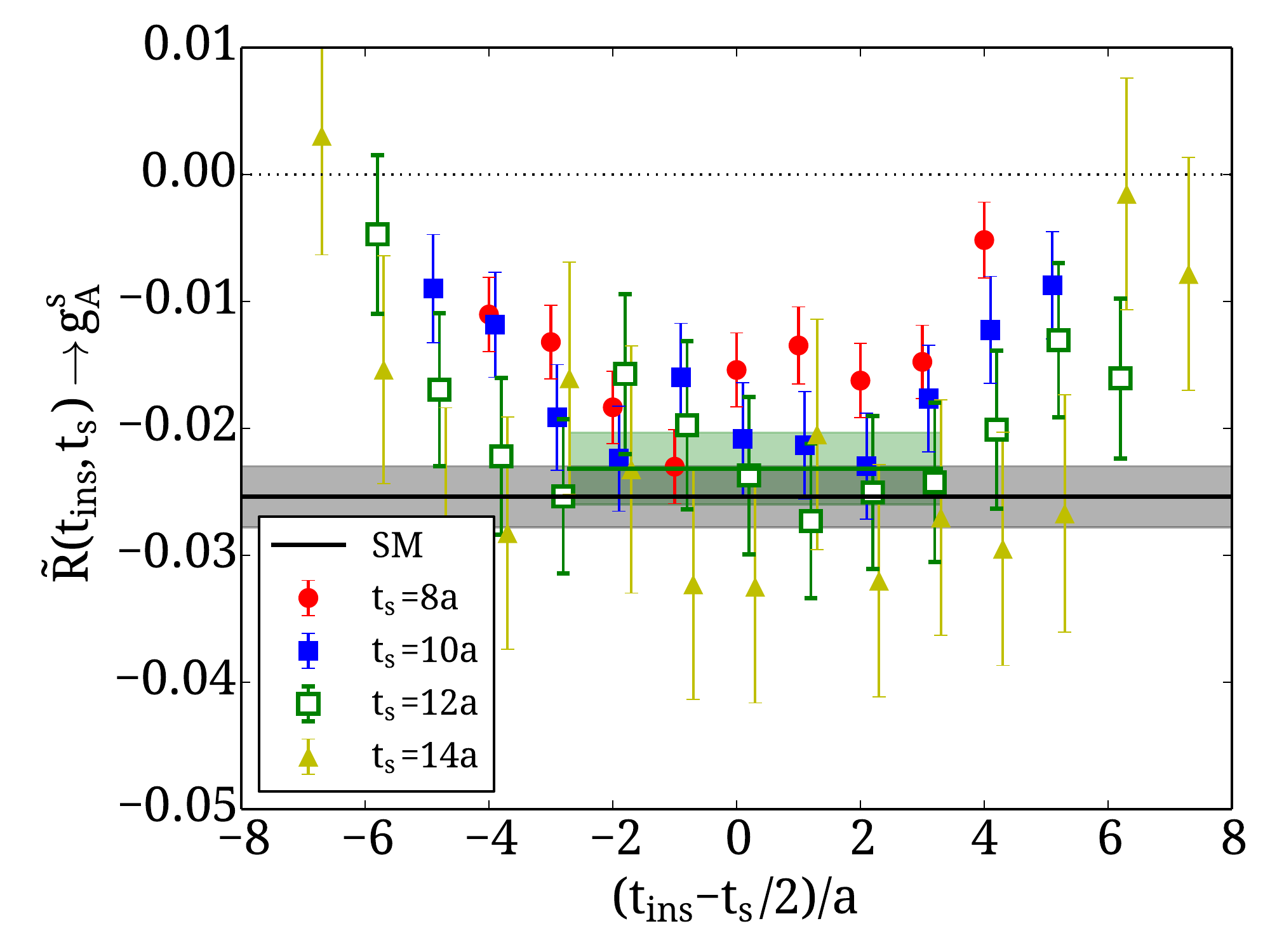}}
\end{minipage}
\caption{Disconnected contributions to the isoscalar (left) and strange (right) nucleon axial charge for the B55 ensemble.}
\label{fig:disconn}
\end{figure}

Disconnected quark loop contributions arise from the coupling of the current to sea quarks. They are notoriously difficult to compute in lattice QCD. 
The technical reason is that one must compute a close quark loop given by
 $L(x)=Tr\left [\Gamma G(x;x)\right]\>$ for a general bilinear ultra-local operator of the form $\bar{\psi}(x)\Gamma\psi(x)$. This  requires the computation of quark propagators from all ${\bf x}$ (all-to-all propagator) and thus it is $L^3$ more expensive as compared to the calculation of hadron masses. The other reason is that
these loops tend to have large gauge noise and therefore large statistics
are necessary to obtain a meaningful result.
Special techniques that utilize stochastic noise  on all spatial lattice sites are utilized in order to allow for the computation of the all-to-all propagator reducing
the number of inversions to  $N_r$   with $N_r\ll L^3$. The gauge noise 
is reduced by increasing statistics at low cost using  low precision inversions and correcting for the bias (truncated solver method (TSM) or all-mode-averaging).
Despite these new approaches the computation of these contributions 
would be too expensive using conventional computers. We  take advantage of graphics cards (GPUs) for which we developed special multi-GPU codes. These computer architectures are ideal for approaches like TSM.
We have illustrated the applicability of these methods by performing a high-statistics analysis using an $N_f=2+1+1$  TMF ensemble with $L=2.6$~fm,  $a$ = 0.082~fm at $m_\pi$ = 373~MeV, referred to as the B55 ensemble. We analyzed 4700 gauge configurations yielding a total of  ${\sim 150,000}$~statistics. 
The results on the disconnected contributions to the nucleon axial charge due to the light quarks  $g^{u+d}_A$ and due to the strange $g^s_A$
are shown in Fig.~\ref{fig:disconn}. We obtain a non-zero negative value, which
is ${\cal O}(10\%)$ for the u- and d-quarks and  has to be taken into account when e.g. discussing
the intrinsic spin $\frac{1}{2}\Delta\Sigma$ carried by quarks in the nucleon.



\subsubsection{Electromagnetic form factors}

  The nucleon electromagnetic form factors are extracted from 

\be \langle N(p',s^\prime) |j^\mu(0) |N (p,s)\rangle  
= \bar u_N (p',s^\prime)  \left[\gamma^\mu { F_1(q^2)}+\frac{i\sigma^{\mu\nu}q_\nu}{2m}{F_2(q^2)} \right]u_N(p,s)\quad.
\ee
We would like to discuss here two studies at near physical pion mass:
the one with the physical point ensemble of ETMC at $m_\pi=130$~MeV~\cite{Koutsou:2014} and the one by LHPC using  $N_f=2+1$ clover fermions configurations produced by the BMW collaboration with $a=0.116$~MeV and  $m_\pi=149$~MeV~\cite{Green:2014xba}.

\begin{figure}[h!]
\begin{minipage}{0.49\linewidth}
{\includegraphics[width=\linewidth]{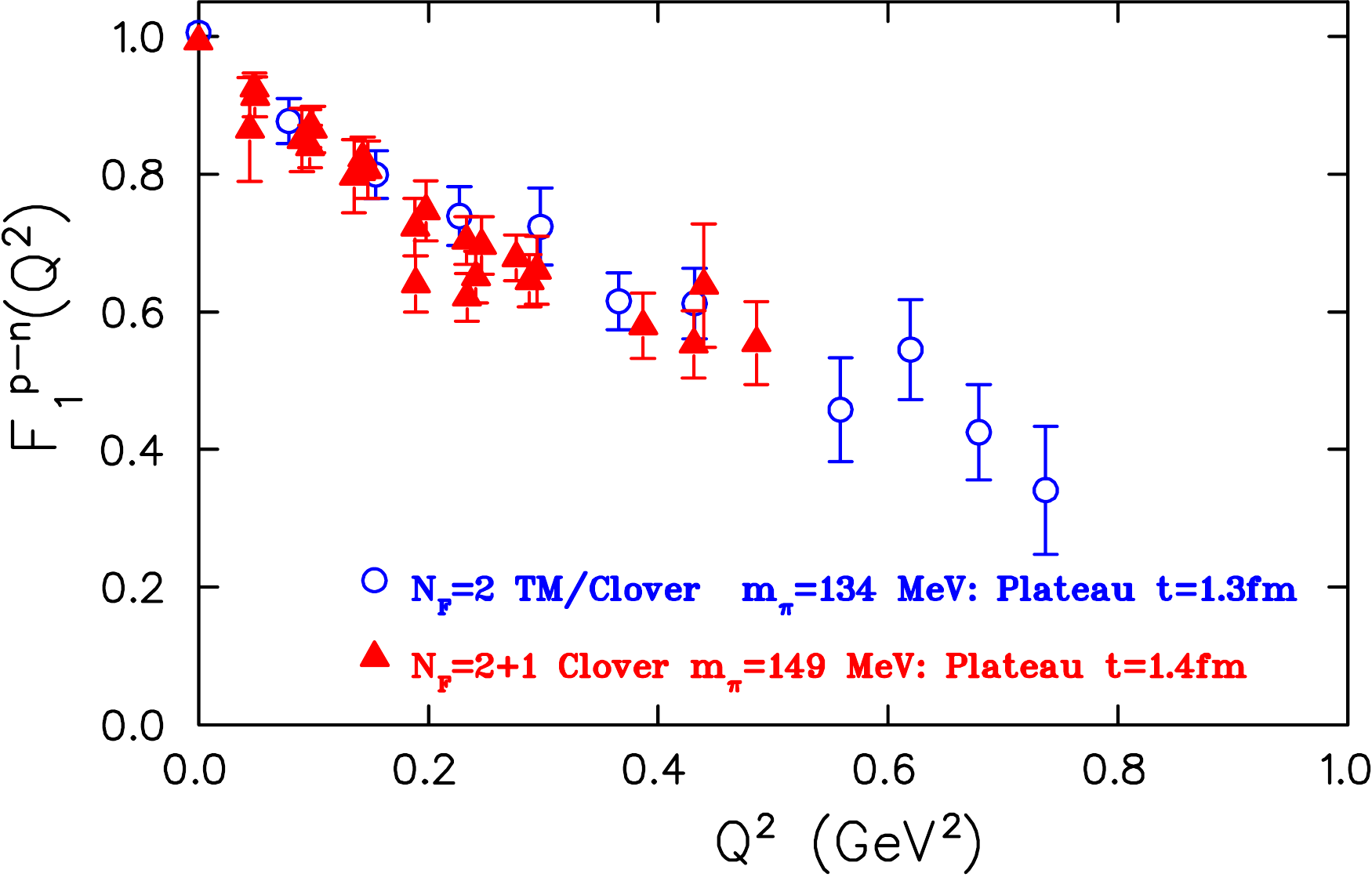}}
\end{minipage}\hfill
\begin{minipage}{0.49\linewidth} \vspace*{0cm}
{\includegraphics[width=\linewidth]{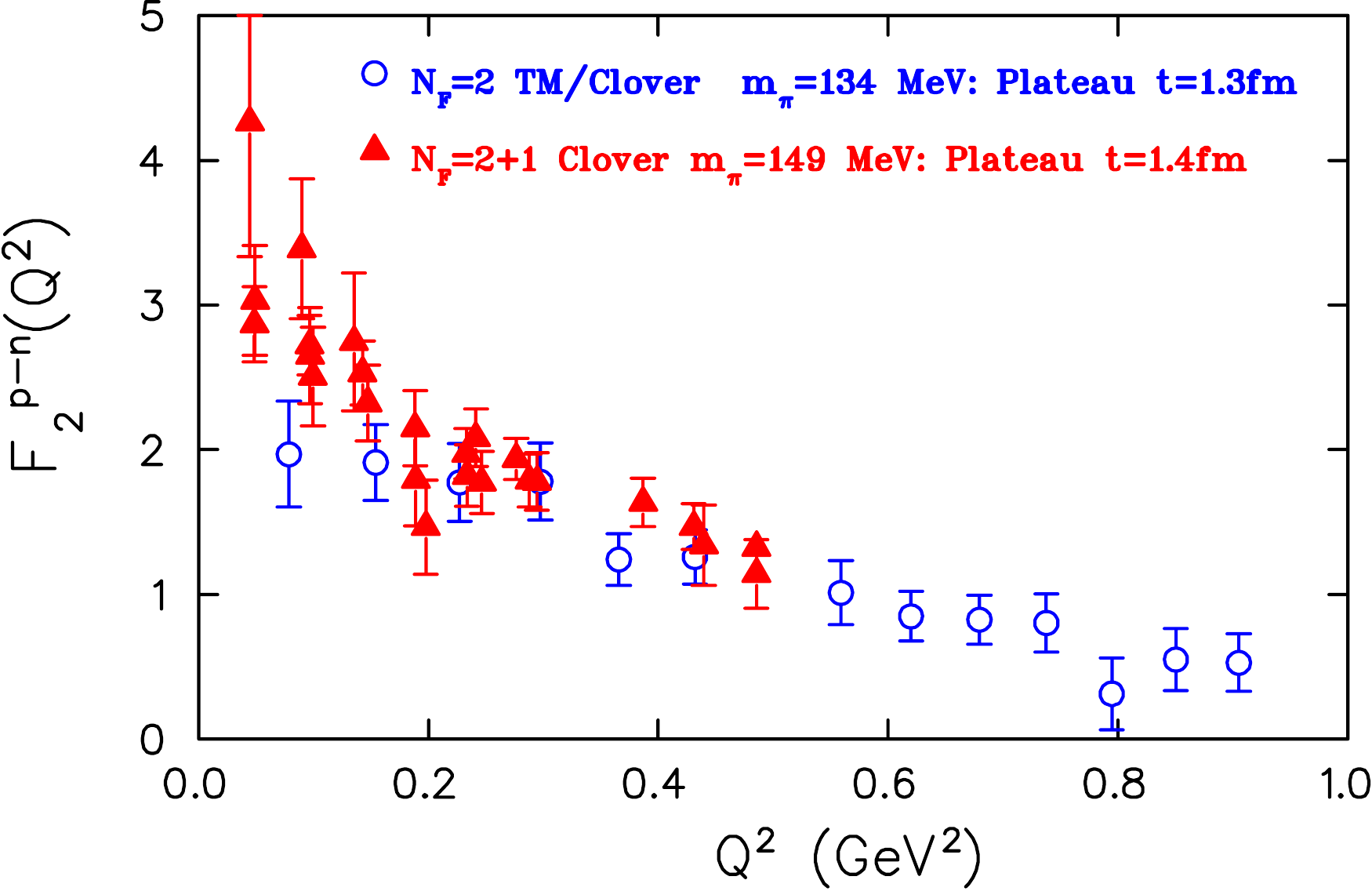}}
\end{minipage}
\caption{Results on the isovector Dirac (left) and Pauli (right) form factors versus $Q^2=-q^2$. The statistics is about $\sim 10^3$ for ETMC and 7750 for LHPC. } 
\label{fig:F1F2}
\end{figure}

\begin{figure}[h!]
\begin{minipage}{0.49\linewidth}\vspace*{-0.5cm}
\hspace*{-0.5cm}{\includegraphics[width=1.1\linewidth,height=1.25\linewidth]{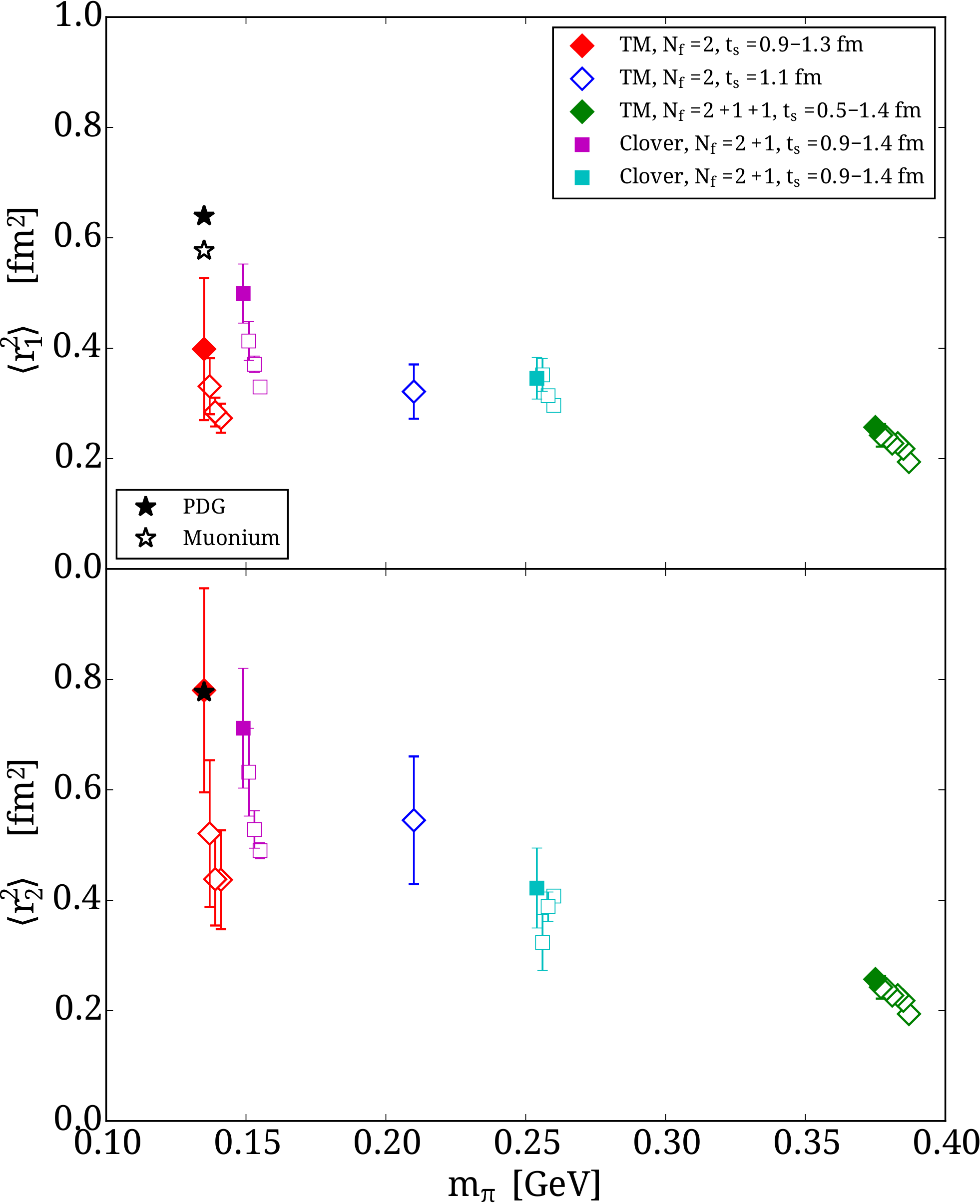}}
\end{minipage}\hfill
\begin{minipage}{0.49\linewidth} \vspace*{0cm}
{\includegraphics[width=\linewidth]{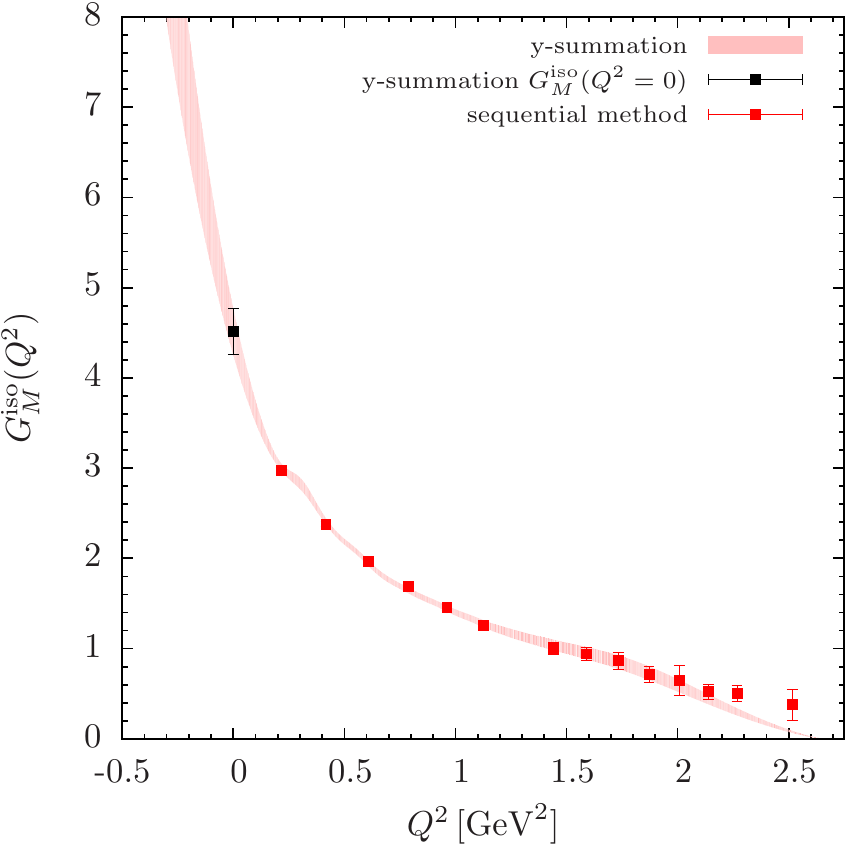}}
\end{minipage}
\caption{Left: Results on the isovector Dirac (top) and Pauli (bottom)  radii
versus $m_\pi^2$ from ETMC (diamonds) and LHPC (squares)~\cite{Koutsou:2014}. Right: Magnetic form factor  directly at $Q^2=0$, as determined by the red band~\cite{Ottnad:2014}. The red points show the results with the conventional method.} 
\label{fig:r1r2}\vspace*{-0.3cm}
\end{figure}
Comparing the results on the Dirac and Pauli form factors between ETMC and LHPC 
in Fig.~\ref{fig:F1F2} we observe an overall agreement independently of the discretization scheme.
The Dirac and Pauli radii can be extracted by fitting the $Q^2$-dependence of $F_1(Q^2)$ and $F_2(Q^2)$ to a dipole form,  $\frac{A_i}{(1+Q^2/M_i^2)^2}$, with $i=1,2, A_1=1$ and taking the derivative: 
 $\langle r_i^2 \rangle =-\frac{6}{F_i}\frac{dF_i}{dQ^2}|_{Q^2=0}=\frac{12}{M_i^2}$.
The results are shown is Fig.~\ref{fig:r1r2} and clearly increase as the pion mass decreases, as well as the sink-source separation increases from $\sim 1$~fm to $\sim 1.3$~fm~(see Ref.~\cite{Koutsou:2014} for more details). However, fitting to a dipole to 
extract the radii introduces a model-dependence. We developed a novel method
that extracts the value directly at $Q^2=0$. The first application of this method was to extract the anomalous magnetic moment determined by the magnetic form factor $G_M(0)$ or equivalently $F_2(0)$. In Fig.~\ref{fig:r1r2}, our results on the isovector $G_M$ for the B55 ensemble are shown  with the red band. As can be seen, the method provides a
good determination of $G_M(0)$ without requiring any assumption of its $Q^2$-dependence~(see Ref.~\cite{Ottnad:2014} for more details).

\section{Conclusions}
 Simulations at the physical point are now feasible and this opens exciting 
possibilities for the study of hadron structure. In this work we presented an overview of
lattice QCD results obtained directly at or close to the physical point
from a number of lattice QCD collaborations, such as results on  the hyperon and charmed baryon masses and isospin splitting, the pion decay constant, as well as, on 
the axial and tensor charges, and electromagnetic form factors  of the nucleon.
We find a value of $g_A$ that is in agreement with experiment and provide a preliminary value for the tensor charge. The computation of disconnected contributions are briefly reviewed focusing on the disconnected quark contributions to
the nucleon axial charge.
First results at the physical point  highlight the need for higher statistics in order that  careful cross-checks can be carried out.
Noise reduction techniques  such as  all-mode-averaging, improved methods for disconnected diagrams and smearing techniques are currently being pursued aiming at  decreasing our errors on the quantities obtained at the physical point.
 When this is achieved, lattice QCD can provide reliable predictions on quantities  probing beyond the standard model physics such as  $g_T$, as well as, on the nucleon $\sigma$-terms.

\section{Acknowledgments}
I would like to thank the members of the ETMC and in particular my close collaborators A. Abdel-Rehim, M. Constantinou, K. Jansen, K. Hadjiyiannakou, C. Kallidonis, G. Koutsou, K. Ottnad, M. Petschlies and A. Vaquero for their contributions to the TMF results shown in this talk.   
Partial support was provided by  the EU ITN project PITN-GA-2009-238353 (ITN STRONGnet) and IA project IA-GA-2011-283286  HadronPhysics3 .
 This work used computational resources provided by PRACE, JSC, Germany and the  Cy-Tera project (NEA Y$\Pi$O$\Delta$OMH/$\Sigma$TPATH/0308/31).

\section{Bibliography}


\begin{footnotesize}



%

\end{footnotesize}


\end{document}